\documentclass[reprint,
 amsmath,amssymb,
 aps,
]{revtex4-1}

\usepackage{graphicx}
\usepackage{dcolumn}
\usepackage{bm}

\begin{document}

\preprint{APS/123-QED}

\title{Modal analysis of noise propagation in a femtosecond oscillator}

\author{Syamsundar De}
\email{syamsundar.de@upb.de}
 \altaffiliation{Integrated Quantum Optics Group, Applied Physics, University of Paderborn, 33098 Paderborn, Germany.}
\author{Val\'erian Thiel}
 \altaffiliation{Clarendon Laboratory, University of Oxford, Parks Road, Oxford, OX1 3PU, UK.}
\author{Jonathan Roslund}
\author{Claude Fabre}
\author{Nicolas Treps}
\affiliation{Laboratoire Kastler Brossel, Sorbonne Universit\'{e}, CNRS, ENS-PSL Research University, Coll\'{e}ge de France, 4 place Jussieu, Paris, 75252, France.}

\date{\today}

\begin{abstract}
We study noise propagation dynamics in a femtosecond oscillator by injecting external noise on the pump intensity. We utilize a spectrally-resolved homodyne detection technique that enables simultaneous measurement of amplitude and phase quadrature noises of different spectral bands of the oscillator. To reveal the impact of added pump noise on the oscillator noise, we perform a modal analysis of the oscillator noise in which each mode corresponds to a particular temporal/spectral shape of the pulsed light. We compare this modal approach with the conventional noise detection methods and find the superiority of our method in particular unveiling a complete physical picture of noise distribution in femtosecond oscillator. 

\end{abstract}
                              
\maketitle

\section{Introduction}
\label{sec:Intro}
Analyzing the noise of a femtosecond oscillator, which features an optical frequency comb (OFC) in the frequency domain, is of paramount importance as its presence limits the sensitivity of OFC-based metrological applications such as spectroscopy, ranging, optical clocks, time/frequency distribution or remote sensing, to name only a few \citep{Udem2002,Marian2004,Diddams2001,Daussy2005,Swann2006}. It turns out that the OFC noise generally manifests itself as fluctuations of only a few global physical parameters such as average power, center-frequency, carrier envelope-offset (CEO) phase, and repetition rate \citep{Haus1990}. Measurement and control of the fluctuations of different physical parameters of OFC are well-established fields of research and multiple methods have already been developed \citep{xu1996route,Telle1999,Diddams2000,Jones2000,Witte2004,Walker2007,Quraishi2014,Coluccellii2015}. However, simultaneous retrieval of the noise from all the parameters using existing methods is not at all straightforward as it usually requires multiple detection schemes and setups, making the whole endeavor cumbersome. Besides, comparing the outcomes of the measurements necessitates a careful calibration of the equipment and normalization. 

In another study \citep{Val2018}, we have introduced a novel measurement technique which allows for the extraction of multiple parameter noise in a single measurement using the same setup. This interferometric scheme combines spectrally-resolved detection using multipixel detectors with balanced homodyne detection, thus permitting simultaneous access of both the amplitude and phase noise of different spectral bands of the OFC. Then to retrieve the fluctuations associated with different comb parameters, a modal analysis is performed in which each OFC parameter relates to a particular noise mode \citep{TheSchmeissnerPaper}.
   
In this study, we follow this modal approach to investigate noise propagation dynamics in an OFC. In particular, we study propagation of pump noise to different physical parameters of the OFC by injecting Gaussian noise on the pump laser intensity. Finally to establish the validity of our modal method, the results are compared with the conventional noise measurement techniques.

\section{Modal description of noise}
\label{sec:theory}
We consider the analytical form of the electric field of a single pulse in the time domain as $E(t) = \mathcal{E}_0 a(t) e^{-i\omega_{0}t}$, where $\mathcal{E}_0$ is the single photon field constant \citep{grynberg2010}, $\omega_0$ is the optical carrier frequency, and $a(t)$ is the slowly-varying complex envelope. Therefore, the complex field of a train of pulses that constitutes an OFC can be expressed as $E_{comb}(t) = \mathcal{E}_0 \sum_{n} a(t-nT) e^{-i\omega_{0}(t-nT)}e^{-in\Delta\phi_{CEO}}$, where $T$ denotes the separation between the pulses and $\Delta\phi_{CEO}$ stands for the carrier envelope-offset (CEO) phase. Consequently in the frequency domain, the OFC field writes as $\widetilde{E}_{comb}(\omega)= \mathcal{E}_0\widetilde{a}(\Omega)\sum_{n}\delta \left[\omega - \left( n\omega_{rep} + \omega_{CEO}\right)\right]$, where '$\sim$' stands for the Fourier component, $\Omega=\omega - \omega_0$ is the optical frequency relative to the carrier, $\omega_{rep} = 2\pi/T$ is the pulse repetition rate, and $\omega_{CEO}=\Delta\phi_{CEO}/T$ is the CEO frequency.  

The presence of different noise sources (mechanical and thermal drifts, spontaneous emission, pump intensity noise) perturbs the coherent comb structure, which can be interpreted in terms of the fluctuations of a single pulse parameters $\vec{p}=(\delta\epsilon, \delta\tau_e, \delta\omega, \delta\tau_c)$ as follows:
    \begin{align}
    E(t,\vec{p}) = \mathcal{E}_0 (1+\delta\epsilon)\, a(t-\delta\tau_e) \, e^{-i (\omega_{0}-\delta\omega)(t-\delta\tau_c)} \label{eq:fieldtime:noise}
    \end{align}
where $\delta\epsilon$, $\delta\tau_e$ denote the fluctuations of pulse envelop amplitude and arrival time, and $\delta\omega$, $\omega_0\delta\tau_c$ represent carrier frequency and phase fluctuations respectively. 
Fourier transform of \eqref{eq:fieldtime:noise} gives the following expression of the fluctuating field in the spectral domain 
    \begin{align}
    \widetilde{E}(\Omega,\vec{p}) = \mathcal{E}_0 (1+\delta\epsilon)\, \widetilde{a}(\Omega-\delta\,\omega) \, e^{i (\omega_{0}\delta\tau_{c}+\Omega\delta\tau_{e})}
    \label{eq:fieldfreq:noise}
    \end{align}
At first order, we can write the field fluctuations $\widetilde{\delta E}(\Omega)=\widetilde{E}(\Omega, \vec{p})-\widetilde{E}(\Omega)$ as	
    \begin{align}
    \widetilde{\delta E}(\Omega) = \mathcal{E}_0 \left[\left(\delta\epsilon + i\omega_0\delta\tau_{\textrm{c}} + i\Omega\delta\tau_{\textrm{e}}\right)\widetilde{a}(\Omega)-\delta\omega\frac{\partial \widetilde{a}(\Omega)}{\partial\Omega}\right]. \label{eq:field:full}
    \end{align}
We write $\widetilde{a}(\Omega)= \alpha(\Omega) \, e^{i \phi}$, where $\alpha (\Omega)$ denotes spectral amplitude and for simplicity we assume a constant spectral phase $\phi$. This leads to the following quadrature description of the field fluctuations:
    \begin{align}
    2\widetilde{\delta E}(\Omega)=\mathcal{E}_0\left[\delta\,x(\Omega)+i\delta\,p(\Omega)\right]e^{i\phi} \label{eq:field:quadratures}
    \end{align}
where amplitude quadrature noise $\delta x(\Omega)=2\delta \alpha(\Omega)$ carries the information of spectral amplitude fluctuation $\delta \alpha(\Omega)$, and phase quadrature noise $\delta p(\Omega)=2 \alpha(\Omega)\delta\phi$ depends on the fluctuation of spectral phase $\delta\phi$. Comparing \eqref{eq:field:full} and \eqref{eq:field:quadratures}, we get 
    \begin{align}
    \delta x(\Omega)&=2\left[\delta\epsilon\alpha(\Omega)-\delta\omega\frac{\partial \alpha(\Omega)}{\partial\Omega}\right]\label{eq:dx},\\
    \delta p(\Omega)&=2 \left(\omega_0\delta\tau_{c}+\Omega\delta\tau_{e}\right)\alpha(\Omega)\label{eq:dp}.
    \end{align}
We now introduce a modal description of the field by considering the spectral envelope as $\alpha(\Omega) = \alpha_0 u (\Omega)$, where $\alpha_0$ is the field spectral amplitude, which is a constant proportional to the square root of the total number of photons contained in the field. $u(\Omega)$ is called the mean-field mode, normalized as $\int d\Omega \left\vert u (\Omega) \right\vert^{2} = 1$, which carries the spectral profile of the comb field. A Gaussian mean-field spectral mode as $u(\Omega)= \frac{1}{\sqrt[1/4]{2\pi\Delta\omega^2}}exp\left(-\frac{\Omega^{2}}{4\Delta\omega^2}\right)$ leads to a modal description of the quadrature noise as follows:
    \begin{align}
    \delta x(\Omega) &= 2 \alpha_0 \left[ \delta\epsilon \cdot u_{amp} (\Omega) +\frac{\delta\omega}{2\Delta\omega} \cdot u_{cent-freq}(\Omega) \right], \label{eq:dx:modal}\\
    \delta p(\Omega) &= 2 \alpha_0 \left[ \omega_0 \delta\tau_{c} \cdot u_{CEO}(\Omega) + \Delta\omega \delta\tau_{e} \cdot u_{rep-rate}(\Omega) \right]\label{eq:dp:modal}
    \end{align}
where $\Delta\omega$ is the spectral bandwidth of the field defined as $\Delta\omega^2 = \int d\Omega\,\Omega^2 \left\vert u (\Omega) \right\vert^{2}$, and $u_k(\Omega)$ is a normalized spectral mode with $k\in\lbrace amp,\,cent-freq,\,CEO,\,rep-rate\rbrace$. On the amplitude quadrature, fluctuations of comb spectral amplitude and center-frequency are respectively attached with amplitude mode $u_{amp} (\Omega) = u(\Omega)$ and the center-frequency mode $u_{cent-freq}(\Omega) = -2 \Delta \omega \cdot \frac{\partial u}{\partial \Omega}$. Likewise, fluctuations of CEO phase and repetition rate lie in the phase quadrature of the field and respectively carried by CEO phase mode $u_{CEO} (\Omega) = u(\Omega)$ and repetition rate mode $u_{rep-rate}(\Omega) =  \frac{\Omega u(\Omega)}{\Delta \omega}$. It is worth noting that $u_{amp} (\Omega) \equiv u_{CEO} (\Omega)$, as well as $u_{cent-freq}(\Omega) \equiv  u_{rep-rate}(\Omega)$ since we consider a Gaussian spectral profile of our OFC. Moreover, it is possible to show that the next high-order mode in amplitude and phase quadrature characterize spectral bandwidth and group-velocity dispersion (GVD) fluctuations respectively \citep{Val2018, jian2012real}. 

Thus, retrieving the fluctuations of different physical parameters requires a measurement scheme that is capable of addressing different quadratures as well as different spectral modes of the comb field. To this aim, we combine homodyne detection, which is a field quadrature-sensitive detection scheme, with spectrally-resolved detection that unfolds the underlying spectral modes of the comb field \citep{Val2018}. 

\section{Methodology}
\label{sec:method}
\begin{figure}[htb]
		\centering
		\includegraphics[width=\linewidth]{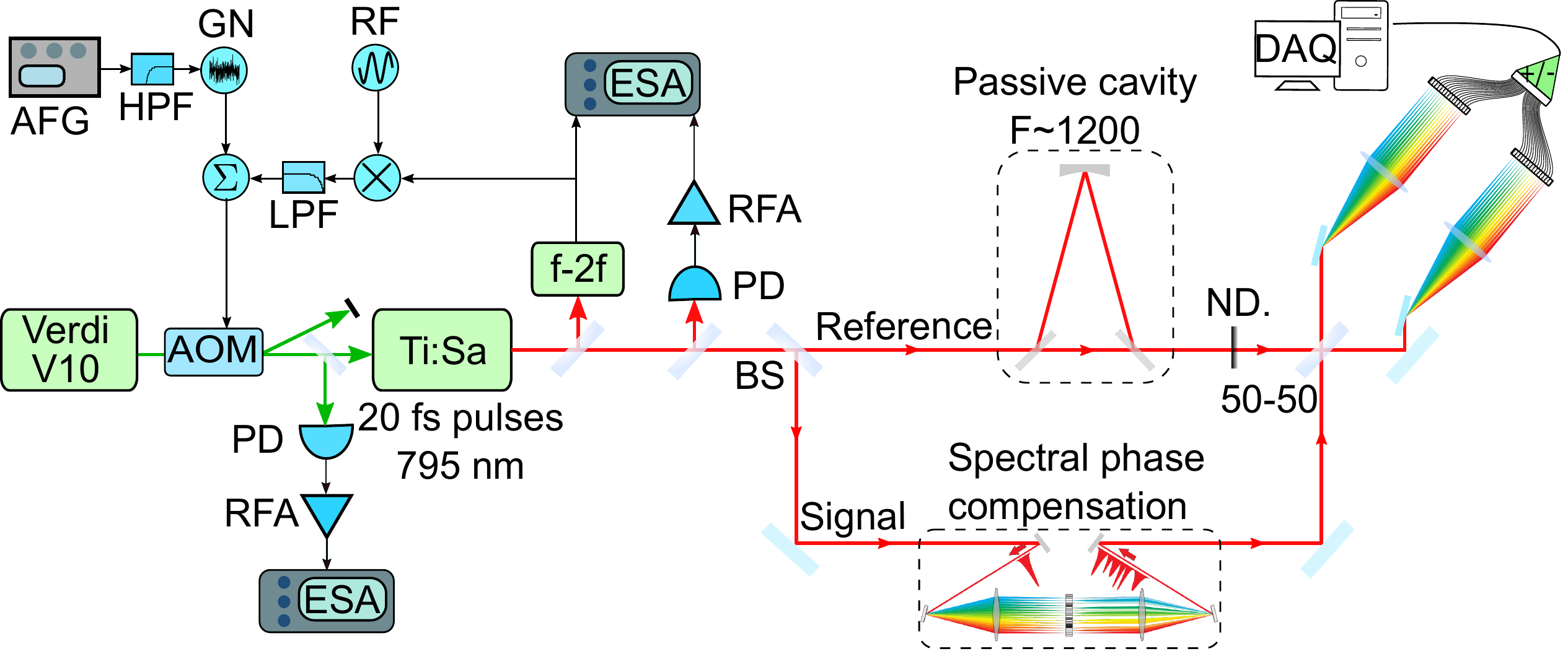}
		\caption{
			General experimental layout. 
			Ti:Sa: titanium-Saphire oscillator; AOM: acousto-optic modulator; f-2f: carrier envelope-offset (CEO) frequency detection using f-2f interferometer scheme; ND: neutral density filter; PD: photo-diode; RFA: radio-frequency amplifier; LPF: low-pass filter; HPF: high-pass filter; RF: radio-frequency signal generator; AFG: arbitrary function generator; GN: Gaussian noise; ESA: electrical spectrum analyzer; DAQ: data acquisition card.
			}
		\label{fig:exp-layout}
	\end{figure}

A general layout of the experimental scheme is depicted in figure \ref{fig:exp-layout}. In this study, we investigate the noise of a commercial Titanium-Sapphire (Ti:Sa) femtosecond oscillator (\textsc{Femtolaser} Synergy) delivering a train of 20 fs pulses at a repetition rate of 156 MHz. The spectrum is well-approximated by a Gaussian distribution of 45 nm full-width-at-half-maximum (FWHM) centered at 795 nm. The oscillator is pumped with a Verdi V-10 (\textsc{Coherent}), set at 5W. The pump beam passes through an acousto-optic modulator (AOM), for which the first order of diffraction is blocked while the fundamental is used to pump the oscillator. This allows us to control the pump power as well as to lock the CEO frequency \cite{Jones2000} to an external high-stability RF source (\textsc{Rhodes-Schwarz} SMA 100 A) using a commercial f-2f interferometer unit (\textsc{Menlosystem}). One part of the f-2f interferometer output is plugged into an electrical spectrum analyzer (ESA) to directly measure the CEO frequency and the repetition rate noise. We also measure the intensity noise of the OFC and the pump laser using conventional setup containing a photodiode (PD) followed by a radio-frequency amplifier (RFA), and finally an electrical spectrum analyzer. To investigate pump noise propagation mechanism, external noise is added to the pump intensity through the AOM by mixing the CEO locking signal with  Gaussian noise from an arbitrary function generator (\textsc{Tektronix} AFG 3022C). This added noise is limited to 2.5 MHz bandwidth due to the sampling rate of the function generator. This noise is high-pass filtered (HPF) at 200 kHz to ensure that it does not perturb the CEO servo-locking of 100 kHz bandwidth.
	
We implement a homodyne detection scheme based on a Mach-Zender interferometer, in which the laser beam is split into a reference and a signal arm \citep{Val2018}. As both fields originate from the same OFC source, a high finesse ($\simeq$ 1200) Fabry-Perot cavity is introduced in the reference arm. It acts on both optical quadratures as a low-pass filter with a cut-off frequency of $f_c \simeq 125$ kHz \cite{TheSchmeissnerPaperAboutCavities}. This effectively decouples both arms of the interferometer, such that the noise measured at the output  predominately originates from the signal beam for analysis frequencies higher than $\simeq 2 \, f_c$ \cite{TheSchmeissnerPaper}. Besides, a pulse-shaper in a $4f-$line is introduced in the signal arm to match the spectral phase between the two arms, thus maximizing the visibility of the interference pattern. The reference beam is adequately attenuated with respect to the signal beam using a variable neutral-density (ND) filter. The reference and the signal are mixed on a 50:50 beamsplitter and subsequently detected with a spectrally-resolved detection scheme at both outputs of the interferometer. Specifically, each output is diffracted with a grating, and then different colors are focused with a microlens array on 8 different pixels of a  photodiode array (S4111-16R Hamamatsu). Each pixel (diode) is connected to a custom-built transimpedance amplifier that splits the DC from the high-frequency part (cut-off frequency of $\simeq 100$ kHz) of the photocurrent. The DC is used for alignment purposes and to measure the comb spectrum $u(\Omega)$ from which the physical modes, shown in Fig.\,\ref{fig:parameter-noise}, are computed. The high-frequency part is subsequently demodulated at a given RF frequency and recorded with an acquisition card. We finally retrieve the noise spectra by post-processing the recorded time traces.
\section{Results}
\label{sec:result}
    \begin{figure}[htb]
    	\centering
    	\includegraphics[width=\linewidth]{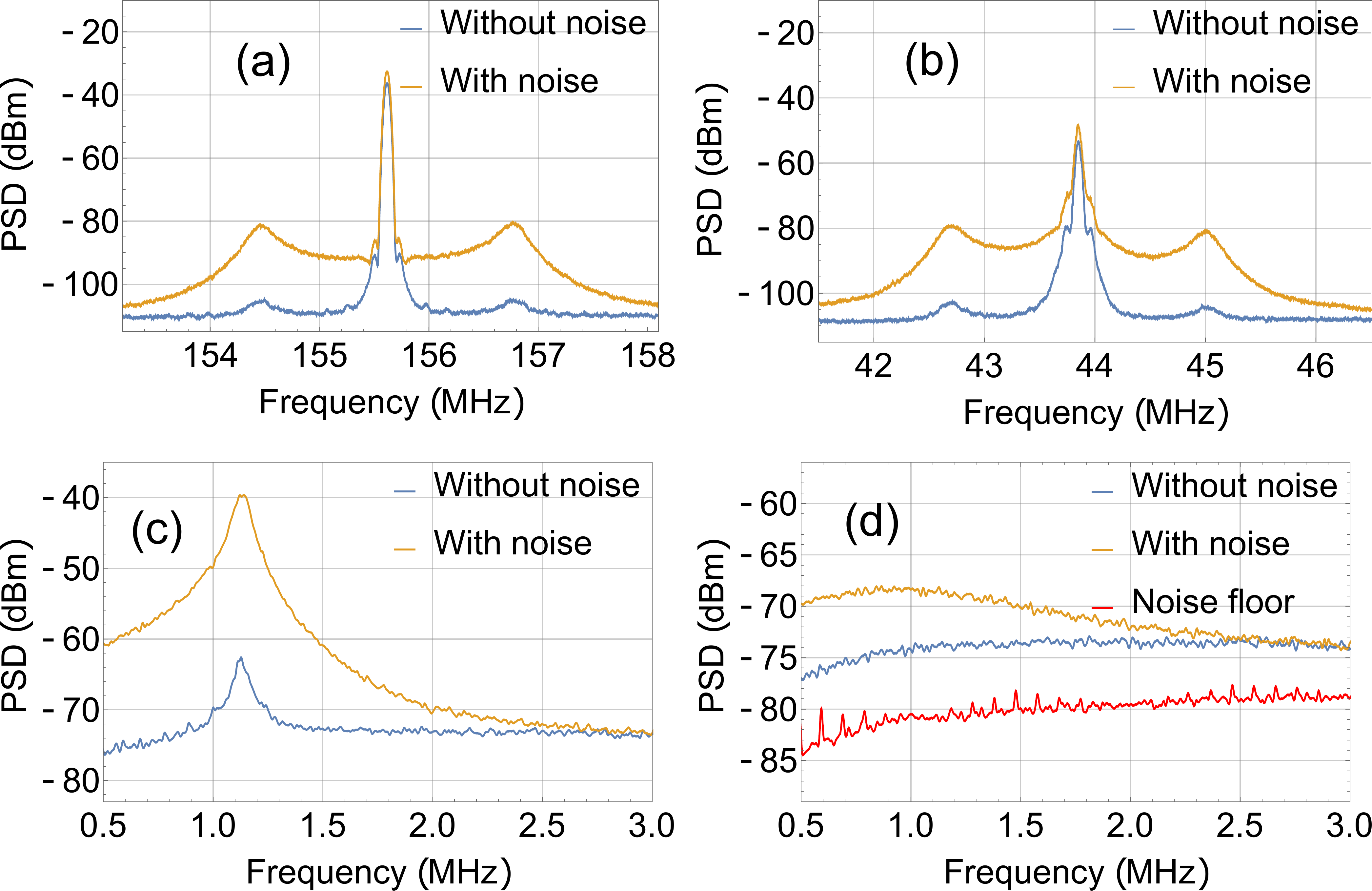}
    	\caption{
    		Power spectral densities (PSDs) of (a) repetition rate, (b) CEO frequency, (c) OFC intensity noise, and (d) pump intensity noise measured with an electrical spectrum analyzer (ESA) for a 40 kHz resolution bandwidth (RBW) and 100 Hz video bandwidth (VBW).
    		}
    	\label{fig:ESA-spectrum}
    \end{figure}
To compare our method with already existing scheme, we first investigate the spectral properties of the fluctuations from different noise sources within sideband frequencies of 0.5 MHz to 3 MHz.
The results of the direct measurement with an ESA are shown in Fig.\,\ref{fig:ESA-spectrum} for (a) repetition rate, (b) CEO frequency, (c) OFC intensity, and (d) pump intensity with and without external pump noise. One can notice that the addition of pump noise increases noise of all the comb parameters. The peak at 1.1 MHz offset frequency in the noise spectra of all the comb parameters originates from the relaxation oscillation mechanism. 
    \begin{figure}[htb]
    	\centering
    	\includegraphics[width=\linewidth]{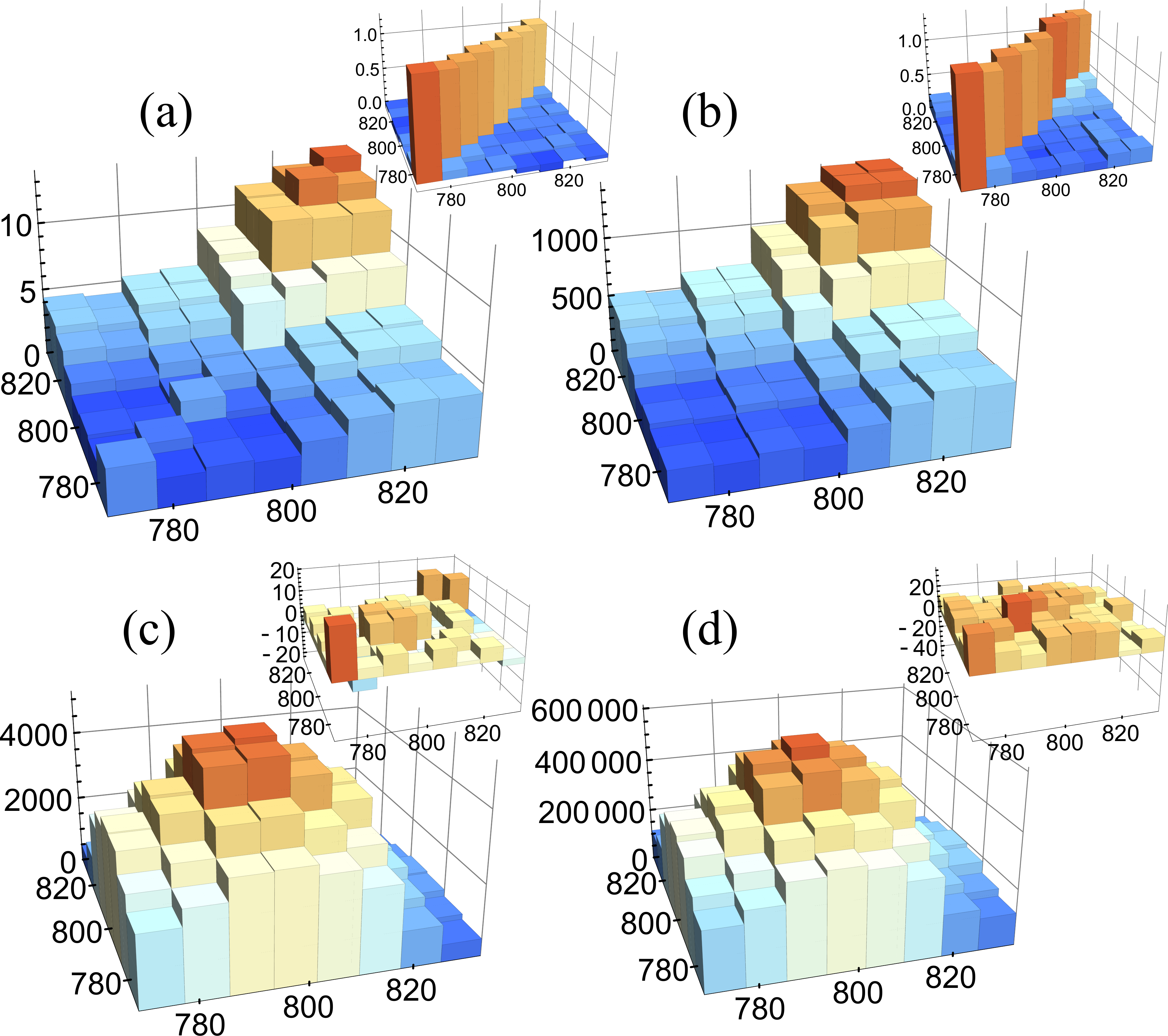}
    	\caption{
    	 Amplitude (a,b) and phase (c,d) quadrature noise covariance matrices respectively without and with  added pump noise. All the noise matrices are normalized to shot noise and correspond to $\sim 1$ MHz  sideband frequency. Insets: Noise matrices at a shot-noise limited sideband frequency ($\sim 3$ MHz). 
    	}
    	\label{fig:noise-matrices}
    \end{figure}
In our modal approach, we rely on the fact that OFC noise follows Gaussian statistics in the range of analyzed sideband frequencies, and hence its behavior can be completely described by a covariance matrix. We therefore compute the spectral covariance matrices for phase $\langle\delta p (\Omega_i)\delta p (\Omega_j)\rangle$ and amplitude quadrature noises $\langle\delta x (\Omega_i)\delta x (\Omega_j)\rangle$, where $\Omega_{n}$ stands for the optical band of the $n-$th pixel in our multi-pixel homodyne scheme. The data analysis to retrieve the noise matrices from the recorded time traces of the multi-pixel homodyne output is detailed in \citep{Val2018}. It is worth mentioning that this method facilitates simultaneous deduction of the amplitude and the phase quadrature noise by taking the sum and the difference of the photocurrents respectively. This simultaneous retrieval of both the quadrature noises enables us to study phase-amplitude noise correlation in OFC as demonstrated in \citep{Val2018}. However, this noise extraction method leads to an imbalance in the two quadrature noise levels that is linked to the optical power difference between the signal and reference field in our experiment. In the data processing, we renormalize the quadrature noises accordingly to take into account this difference in optical power and to compare them on the same scale. 

In Fig.\,\ref{fig:noise-matrices}, we show the covariance matrices for a sideband frequency of $\sim 1$ MHz at which the OFC dynamics is important. Additionally, the insets display the noise matrices for a sideband frequency of $\sim 3$ MHz, where OFC noise reaches shot noise limit. The noise matrices are normalized to shot noise in all the cases. The left column corresponds to no external pump noise, and in this case the characteristic behavior of the noise matrices at different sideband frequencies have been thoroughly discussed in \citep{Val2018}. In particular, we remind that the correlations between the noise of different optical bands at low sideband frequencies are linked to the mode-locking mechanism, whereas these spectral noise correlations disappear at shot noise limited sideband frequencies. Nevertheless, the main emphasis here is to study the impact of external pump noise on the comb noise dynamics, and the corresponding noise matrices are depicted along the right column of Fig.\,\ref{fig:noise-matrices}. One can notice that adding external noise to the pump does not really alter the overall structure of the noise matrices, except from the global increment of their magnitudes. Similar trend is also observed for phase-amplitude noise correlation matrices which are not shown here. This indicates  that the spectral noise correlations within the analyzed sideband frequencies are predominantly governed by mode-locking dynamics of the OFC.

    \begin{figure}[htb]
    	\centering
    	\includegraphics[width=0.7\linewidth]{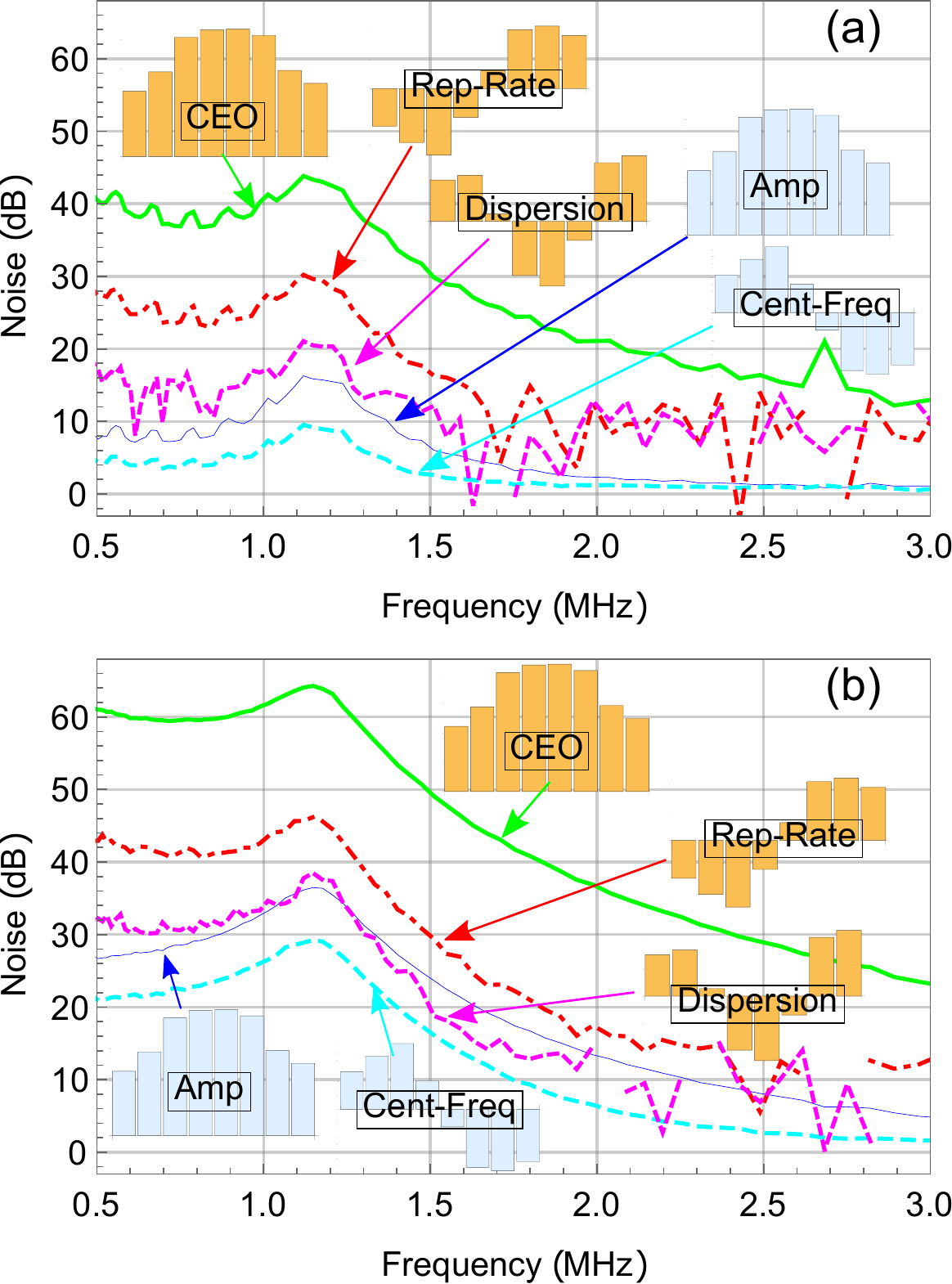}
    	\caption{
    		Noise spectra of the comb parameters obtained by decomposing the covariance matrices in the physical mode basis. (a) Without and (b) with  external noise added to the pump.
    	}
    	\label{fig:parameter-noise}
    \end{figure}
 To reveal the underlying physical process behind OFC noise dynamics, one approach is eigendecomposing  the noise matrices and finding the independent noise modes as introduced in \citep{TheSchmeissnerPaper}. This provides three significant phase modes and two dominant amplitude modes in our OFC that reasonably reproduces the theoretical modes \citep{Haus1990}.  However, for a better physical understanding we analyze the noise matrices in the basis of physical modes associated with the global parameters of the OFC as described in \eqref{eq:dx:modal}, \eqref{eq:dp:modal}. This unveils the noise distribution in different global parameters of the comb as shown in Fig.\,\ref{fig:parameter-noise} for the two cases, (a) without and (b) with external pump noise. The spectra of the parameter noises for no external pump noise have already been reported in \citep{Val2018}. Specifically, there are mainly three phase modes and two amplitude modes, and among all the strongest contribution comes from CEO phase fluctuations. The other two important phase noise mechanisms are repetition rate and  intra-cavity group-velocity dispersion (GVD) fluctuations. On the amplitude quadrature, contributions from power and center-frequency fluctuations are significant. Moreover, all the noise spectra exhibit relaxation oscillation at about 1.1 MHz and  reach the shot noise limit at high sideband frequencies due to low-pass filtering effect of the oscillator cavity. However, the main emphasis here is finding the impact of adding external noise to the pump, and Fig.\,\ref{fig:parameter-noise} demonstrates that the spectral behavior of the parameter noises as well as their relative strengths remain unchanged by the excess pump noise, although overall noise levels of all the global parameters are increased. This insensitivity of spectral behavior of parameter noises on exact pump noise levels reiterates the importance of mode-locking mechanism behind the   noise correlations of different optical bands of the OFC in accordance with the results of Fig.\,\ref{fig:noise-matrices}. Furthermore, our modal method shows increment of noise levels in all the parameters due to pump noise augmentation, thus reproduces the noise propagation mechanism in our OFC as found in direct measurement, Fig.\,\ref{fig:ESA-spectrum}. One possible explanation for the pump noise propagation reads as: the pump noise creates gain fluctuations, which can give rise to  power and center-frequency fluctuations and that might eventually trigger fluctuations of CEO phase, timing, dispersion via phase-intensity coupling driven by Kerr-lens modelocking mechanism \citep{xu1996route,holman2003detailed,Helbing2002}. 
 
\begin{figure}[htbp]
    	\centering
    	\includegraphics[width=0.7\linewidth]{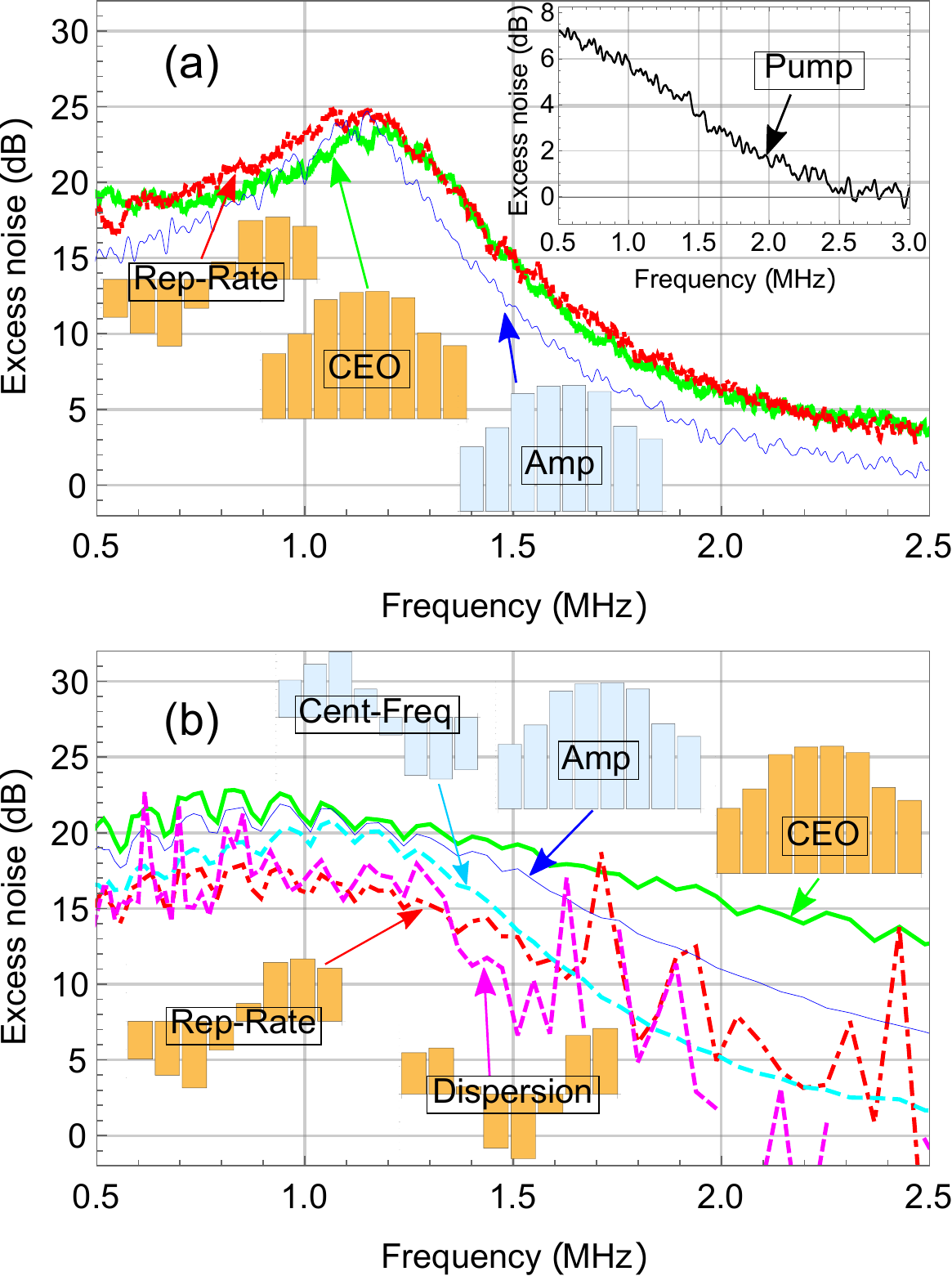}
    	\caption{
    		Excess noise obtained from (a) direct measurement with an ESA (inset: excess pump noise), and (b) modal method.
    		}
    		\label{fig:excess-noise}
    \end{figure}

Finally, studying pump noise propagation also facilitates to compare our modal method with the direct approach employing ESA. In this regard, we compute the difference between the two cases, with and without external pump noise, for each parameter noise. We name this as the excess noise of each physical parameter, and Fig.\,\ref{fig:excess-noise}
delineates the excess noise spectra measured with the two aforementioned techniques. Both the approaches reproduce quite similar noise spectra, which on one hand establishes the validity of our modal approach. On the other hand, this as well demonstrates the superiority of the modal method over the commonly employed direct approach. In particular, while direct approach does not permit straight forward measurement of center-frequency, dispersion noise in addition to the fact that each parameter noise measurement requires a dedicated setup, the modal method provides a complete picture of noise distribution in all the physical parameters of the OFC using a single setup. It is important to mention that the mismatch between the direct and modal method, particularly for the phase parameters at shot noise limited frequencies, comes from the noise added by the renormalization which can be ignored.

\section{Conclusion}
\label{sec:conclusion}
To conclude, we present a novel study that provides a complete knowledge of pump noise propagation in all the global physical parameters of an OFC in a single measurement. Our method is based on a temporal/spectral mode-dependent noise analysis, which we accomplish by developing a setup that combines spectrally-resolved detection with homodyne detection. This scheme enables us to measure all the phase and amplitude noises along with their correlation in a single measurement. In this study, we demonstrate pump noise propagation by adding external noise to the pump intensity. For deeper understanding of OFC noise dynamics, it would be interesting to investigate the impact of noise injections, for example, on the cavity length by acting on the piezo of the output coupler, or on the intra-cavity dispersion with the help of a piezo-mounted wedge inside the cavity, etc. This would help to figure out different sources of noises for the OFC and eventually to design an optimal feedback loop to simultaneously control noises of multiple OFC parameters. It is worth mentioning that this method is not limited to Ti:Sa OFCs, and may be applied to analyze noises of other sources such as fiber \citep{Newbury2007} and micro-resonator based OFCs \citep{Herr2014}.

\section*{Funding Information}

The work was supported by the European Research Council starting grant Frecquam, by the French National Research Agency project SPOCQ and LASAGNE, and received funding from the European Union's Horizon 2020 research and innovation programme under Grant Agreement No. 665148. J. R. acknowledges support from the European Commission through Marie Curie Actions QOCO.

\bibliography{Manuscript}

\begin{thebibliography}{23}%
\makeatletter
\providecommand \@ifxundefined [1]{%
 \@ifx{#1\undefined}
}%
\providecommand \@ifnum [1]{%
 \ifnum #1\expandafter \@firstoftwo
 \else \expandafter \@secondoftwo
 \fi
}%
\providecommand \@ifx [1]{%
 \ifx #1\expandafter \@firstoftwo
 \else \expandafter \@secondoftwo
 \fi
}%
\providecommand \natexlab [1]{#1}%
\providecommand \enquote  [1]{``#1''}%
\providecommand \bibnamefont  [1]{#1}%
\providecommand \bibfnamefont [1]{#1}%
\providecommand \citenamefont [1]{#1}%
\providecommand \href@noop [0]{\@secondoftwo}%
\providecommand \href [0]{\begingroup \@sanitize@url \@href}%
\providecommand \@href[1]{\@@startlink{#1}\@@href}%
\providecommand \@@href[1]{\endgroup#1\@@endlink}%
\providecommand \@sanitize@url [0]{\catcode `\\12\catcode `\$12\catcode
  `\&12\catcode `\#12\catcode `\^12\catcode `\_12\catcode `\%12\relax}%
\providecommand \@@startlink[1]{}%
\providecommand \@@endlink[0]{}%
\providecommand \url  [0]{\begingroup\@sanitize@url \@url }%
\providecommand \@url [1]{\endgroup\@href {#1}{\urlprefix }}%
\providecommand \urlprefix  [0]{URL }%
\providecommand \Eprint [0]{\href }%
\providecommand \doibase [0]{http://dx.doi.org/}%
\providecommand \selectlanguage [0]{\@gobble}%
\providecommand \bibinfo  [0]{\@secondoftwo}%
\providecommand \bibfield  [0]{\@secondoftwo}%
\providecommand \translation [1]{[#1]}%
\providecommand \BibitemOpen [0]{}%
\providecommand \bibitemStop [0]{}%
\providecommand \bibitemNoStop [0]{.\EOS\space}%
\providecommand \EOS [0]{\spacefactor3000\relax}%
\providecommand \BibitemShut  [1]{\csname bibitem#1\endcsname}%
\let\auto@bib@innerbib\@empty
\bibitem [{\citenamefont {Udem}\ \emph {et~al.}(2002)\citenamefont {Udem},
  \citenamefont {Holzwarth},\ and\ \citenamefont {H\"{a}nsch}}]{Udem2002}%
  \BibitemOpen
  \bibfield  {author} {\bibinfo {author} {\bibfnamefont {T.}~\bibnamefont
  {Udem}}, \bibinfo {author} {\bibfnamefont {R.}~\bibnamefont {Holzwarth}}, \
  and\ \bibinfo {author} {\bibfnamefont {T.~K.}\ \bibnamefont {H\"{a}nsch}},\
  }\href@noop {} {\bibfield  {journal} {\bibinfo  {journal} {Nature}\ }\textbf
  {\bibinfo {volume} {416}},\ \bibinfo {pages} {233} (\bibinfo {year}
  {2002})}\BibitemShut {NoStop}%
\bibitem [{\citenamefont {Marian}\ \emph {et~al.}(2004)\citenamefont {Marian},
  \citenamefont {Stowe}, \citenamefont {Lawall}, \citenamefont {Felinto},\ and\
  \citenamefont {Ye}}]{Marian2004}%
  \BibitemOpen
  \bibfield  {author} {\bibinfo {author} {\bibfnamefont {A.}~\bibnamefont
  {Marian}}, \bibinfo {author} {\bibfnamefont {M.~C.}\ \bibnamefont {Stowe}},
  \bibinfo {author} {\bibfnamefont {J.~R.}\ \bibnamefont {Lawall}}, \bibinfo
  {author} {\bibfnamefont {D.}~\bibnamefont {Felinto}}, \ and\ \bibinfo
  {author} {\bibfnamefont {J.}~\bibnamefont {Ye}},\ }\href@noop {} {\bibfield
  {journal} {\bibinfo  {journal} {Science}\ }\textbf {\bibinfo {volume}
  {306}},\ \bibinfo {pages} {2063} (\bibinfo {year} {2004})}\BibitemShut
  {NoStop}%
\bibitem [{\citenamefont {Diddams}\ \emph {et~al.}(2001)\citenamefont
  {Diddams}, \citenamefont {Udem}, \citenamefont {Bergquist}, \citenamefont
  {Curtis}, \citenamefont {Drullinger}, \citenamefont {Hollberg}, \citenamefont
  {Itano}, \citenamefont {Lee}, \citenamefont {Oates}, \citenamefont {Vogel},\
  and\ \citenamefont {Wineland}}]{Diddams2001}%
  \BibitemOpen
  \bibfield  {author} {\bibinfo {author} {\bibfnamefont {S.~A.}\ \bibnamefont
  {Diddams}}, \bibinfo {author} {\bibfnamefont {T.}~\bibnamefont {Udem}},
  \bibinfo {author} {\bibfnamefont {J.~C.}\ \bibnamefont {Bergquist}}, \bibinfo
  {author} {\bibfnamefont {E.~A.}\ \bibnamefont {Curtis}}, \bibinfo {author}
  {\bibfnamefont {R.~E.}\ \bibnamefont {Drullinger}}, \bibinfo {author}
  {\bibfnamefont {L.}~\bibnamefont {Hollberg}}, \bibinfo {author}
  {\bibfnamefont {W.~M.}\ \bibnamefont {Itano}}, \bibinfo {author}
  {\bibfnamefont {W.~D.}\ \bibnamefont {Lee}}, \bibinfo {author} {\bibfnamefont
  {C.~W.}\ \bibnamefont {Oates}}, \bibinfo {author} {\bibfnamefont {K.~R.}\
  \bibnamefont {Vogel}}, \ and\ \bibinfo {author} {\bibfnamefont {D.~J.}\
  \bibnamefont {Wineland}},\ }\href@noop {} {\bibfield  {journal} {\bibinfo
  {journal} {Science}\ }\textbf {\bibinfo {volume} {293}},\ \bibinfo {pages}
  {825} (\bibinfo {year} {2001})}\BibitemShut {NoStop}%
\bibitem [{\citenamefont {Daussy}\ \emph {et~al.}(2005)\citenamefont {Daussy},
  \citenamefont {Lopez}, \citenamefont {Amy-Klein}, \citenamefont {Goncharov},
  \citenamefont {Guinet}, \citenamefont {Chardonnet}, \citenamefont
  {Narbonneau}, \citenamefont {Lours}, \citenamefont {Chambon}, \citenamefont
  {Bize}, \citenamefont {Clairon}, \citenamefont {Santarelli}, \citenamefont
  {Tobar},\ and\ \citenamefont {Luiten}}]{Daussy2005}%
  \BibitemOpen
  \bibfield  {author} {\bibinfo {author} {\bibfnamefont {C.}~\bibnamefont
  {Daussy}}, \bibinfo {author} {\bibfnamefont {O.}~\bibnamefont {Lopez}},
  \bibinfo {author} {\bibfnamefont {A.}~\bibnamefont {Amy-Klein}}, \bibinfo
  {author} {\bibfnamefont {A.}~\bibnamefont {Goncharov}}, \bibinfo {author}
  {\bibfnamefont {M.}~\bibnamefont {Guinet}}, \bibinfo {author} {\bibfnamefont
  {C.}~\bibnamefont {Chardonnet}}, \bibinfo {author} {\bibfnamefont
  {F.}~\bibnamefont {Narbonneau}}, \bibinfo {author} {\bibfnamefont
  {M.}~\bibnamefont {Lours}}, \bibinfo {author} {\bibfnamefont
  {D.}~\bibnamefont {Chambon}}, \bibinfo {author} {\bibfnamefont
  {S.}~\bibnamefont {Bize}}, \bibinfo {author} {\bibfnamefont {A.}~\bibnamefont
  {Clairon}}, \bibinfo {author} {\bibfnamefont {G.}~\bibnamefont {Santarelli}},
  \bibinfo {author} {\bibfnamefont {M.~E.}\ \bibnamefont {Tobar}}, \ and\
  \bibinfo {author} {\bibfnamefont {A.~N.}\ \bibnamefont {Luiten}},\
  }\href@noop {} {\bibfield  {journal} {\bibinfo  {journal} {Physical Review
  Letters}\ }\textbf {\bibinfo {volume} {94}},\ \bibinfo {pages} {203904}
  (\bibinfo {year} {2005})}\BibitemShut {NoStop}%
\bibitem [{\citenamefont {Swann}\ and\ \citenamefont
  {Newbury}(2006)}]{Swann2006}%
  \BibitemOpen
  \bibfield  {author} {\bibinfo {author} {\bibfnamefont {W.~C.}\ \bibnamefont
  {Swann}}\ and\ \bibinfo {author} {\bibfnamefont {N.~R.}\ \bibnamefont
  {Newbury}},\ }\href@noop {} {\bibfield  {journal} {\bibinfo  {journal}
  {Optics Letters}\ }\textbf {\bibinfo {volume} {31}},\ \bibinfo {pages} {826}
  (\bibinfo {year} {2006})}\BibitemShut {NoStop}%
\bibitem [{\citenamefont {Haus}\ and\ \citenamefont {Lai}(1990)}]{Haus1990}%
  \BibitemOpen
  \bibfield  {author} {\bibinfo {author} {\bibfnamefont {H.~A.}\ \bibnamefont
  {Haus}}\ and\ \bibinfo {author} {\bibfnamefont {Y.}~\bibnamefont {Lai}},\
  }\href {\doibase 10.1364/JOSAB.7.000386} {\bibfield  {journal} {\bibinfo
  {journal} {Journal of the Optical Society of America B}\ }\textbf {\bibinfo
  {volume} {7}},\ \bibinfo {pages} {386} (\bibinfo {year} {1990})}\BibitemShut
  {NoStop}%
\bibitem [{\citenamefont {Xu}\ \emph {et~al.}(1996)\citenamefont {Xu},
  \citenamefont {H{\"a}nsch}, \citenamefont {Spielmann}, \citenamefont {Poppe},
  \citenamefont {Brabec},\ and\ \citenamefont {Krausz}}]{xu1996route}%
  \BibitemOpen
  \bibfield  {author} {\bibinfo {author} {\bibfnamefont {L.}~\bibnamefont
  {Xu}}, \bibinfo {author} {\bibfnamefont {T.~W.}\ \bibnamefont {H{\"a}nsch}},
  \bibinfo {author} {\bibfnamefont {C.}~\bibnamefont {Spielmann}}, \bibinfo
  {author} {\bibfnamefont {A.}~\bibnamefont {Poppe}}, \bibinfo {author}
  {\bibfnamefont {T.}~\bibnamefont {Brabec}}, \ and\ \bibinfo {author}
  {\bibfnamefont {F.}~\bibnamefont {Krausz}},\ }\href@noop {} {\bibfield
  {journal} {\bibinfo  {journal} {Optics Letters}\ }\textbf {\bibinfo {volume}
  {21}},\ \bibinfo {pages} {2008} (\bibinfo {year} {1996})}\BibitemShut
  {NoStop}%
\bibitem [{\citenamefont {Telle}\ \emph {et~al.}(1999)\citenamefont {Telle},
  \citenamefont {Steinmeyer}, \citenamefont {Dunlop}, \citenamefont {Stenger},
  \citenamefont {Sutter},\ and\ \citenamefont {Keller}}]{Telle1999}%
  \BibitemOpen
  \bibfield  {author} {\bibinfo {author} {\bibfnamefont {H.~R.}\ \bibnamefont
  {Telle}}, \bibinfo {author} {\bibfnamefont {G.}~\bibnamefont {Steinmeyer}},
  \bibinfo {author} {\bibfnamefont {A.~E.}\ \bibnamefont {Dunlop}}, \bibinfo
  {author} {\bibfnamefont {J.}~\bibnamefont {Stenger}}, \bibinfo {author}
  {\bibfnamefont {D.~H.}\ \bibnamefont {Sutter}}, \ and\ \bibinfo {author}
  {\bibfnamefont {U.}~\bibnamefont {Keller}},\ }\href@noop {} {\bibfield
  {journal} {\bibinfo  {journal} {Applied Physics B}\ }\textbf {\bibinfo
  {volume} {69}},\ \bibinfo {pages} {327} (\bibinfo {year} {1999})}\BibitemShut
  {NoStop}%
\bibitem [{\citenamefont {Diddams}\ \emph {et~al.}(2000)\citenamefont
  {Diddams}, \citenamefont {Jones}, \citenamefont {Ye}, \citenamefont
  {Cundiff}, \citenamefont {Hall}, \citenamefont {Ranka}, \citenamefont
  {Windeler}, \citenamefont {Holzwarth}, \citenamefont {Udem},\ and\
  \citenamefont {H{\"a}nsch}}]{Diddams2000}%
  \BibitemOpen
  \bibfield  {author} {\bibinfo {author} {\bibfnamefont {S.~A.}\ \bibnamefont
  {Diddams}}, \bibinfo {author} {\bibfnamefont {D.~J.}\ \bibnamefont {Jones}},
  \bibinfo {author} {\bibfnamefont {J.}~\bibnamefont {Ye}}, \bibinfo {author}
  {\bibfnamefont {S.~T.}\ \bibnamefont {Cundiff}}, \bibinfo {author}
  {\bibfnamefont {J.~L.}\ \bibnamefont {Hall}}, \bibinfo {author}
  {\bibfnamefont {J.~K.}\ \bibnamefont {Ranka}}, \bibinfo {author}
  {\bibfnamefont {R.~S.}\ \bibnamefont {Windeler}}, \bibinfo {author}
  {\bibfnamefont {R.}~\bibnamefont {Holzwarth}}, \bibinfo {author}
  {\bibfnamefont {T.}~\bibnamefont {Udem}}, \ and\ \bibinfo {author}
  {\bibfnamefont {T.~W.}\ \bibnamefont {H{\"a}nsch}},\ }\href@noop {}
  {\bibfield  {journal} {\bibinfo  {journal} {Physical Review Letters}\
  }\textbf {\bibinfo {volume} {84}},\ \bibinfo {pages} {5102} (\bibinfo {year}
  {2000})}\BibitemShut {NoStop}%
\bibitem [{\citenamefont {Jones}\ \emph {et~al.}(2000)\citenamefont {Jones},
  \citenamefont {Diddams}, \citenamefont {Ranka}, \citenamefont {Stentz},
  \citenamefont {Windeler}, \citenamefont {Hall},\ and\ \citenamefont
  {Cundiff}}]{Jones2000}%
  \BibitemOpen
  \bibfield  {author} {\bibinfo {author} {\bibfnamefont {D.~J.}\ \bibnamefont
  {Jones}}, \bibinfo {author} {\bibfnamefont {S.~A.}\ \bibnamefont {Diddams}},
  \bibinfo {author} {\bibfnamefont {J.~K.}\ \bibnamefont {Ranka}}, \bibinfo
  {author} {\bibfnamefont {A.}~\bibnamefont {Stentz}}, \bibinfo {author}
  {\bibfnamefont {R.~S.}\ \bibnamefont {Windeler}}, \bibinfo {author}
  {\bibfnamefont {J.~L.}\ \bibnamefont {Hall}}, \ and\ \bibinfo {author}
  {\bibfnamefont {S.~T.}\ \bibnamefont {Cundiff}},\ }\href@noop {} {\bibfield
  {journal} {\bibinfo  {journal} {Science}\ }\textbf {\bibinfo {volume}
  {288}},\ \bibinfo {pages} {635} (\bibinfo {year} {2000})}\BibitemShut
  {NoStop}%
\bibitem [{\citenamefont {Witte}\ \emph {et~al.}(2004)\citenamefont {Witte},
  \citenamefont {Zinkstok}, \citenamefont {Hogervorst},\ and\ \citenamefont
  {Eikema}}]{Witte2004}%
  \BibitemOpen
  \bibfield  {author} {\bibinfo {author} {\bibfnamefont {S.}~\bibnamefont
  {Witte}}, \bibinfo {author} {\bibfnamefont {R.~T.}\ \bibnamefont {Zinkstok}},
  \bibinfo {author} {\bibfnamefont {W.}~\bibnamefont {Hogervorst}}, \ and\
  \bibinfo {author} {\bibfnamefont {K.~S.~E.}\ \bibnamefont {Eikema}},\
  }\href@noop {} {\bibfield  {journal} {\bibinfo  {journal} {Applied Physics
  B}\ }\textbf {\bibinfo {volume} {78}},\ \bibinfo {pages} {5} (\bibinfo {year}
  {2004})}\BibitemShut {NoStop}%
\bibitem [{\citenamefont {Walker}\ \emph {et~al.}(2007)\citenamefont {Walker},
  \citenamefont {Udem}, \citenamefont {Gohle}, \citenamefont {Stein},\ and\
  \citenamefont {H{\''}ansch}}]{Walker2007}%
  \BibitemOpen
  \bibfield  {author} {\bibinfo {author} {\bibfnamefont {D.~R.}\ \bibnamefont
  {Walker}}, \bibinfo {author} {\bibfnamefont {T.}~\bibnamefont {Udem}},
  \bibinfo {author} {\bibfnamefont {C.}~\bibnamefont {Gohle}}, \bibinfo
  {author} {\bibfnamefont {B.}~\bibnamefont {Stein}}, \ and\ \bibinfo {author}
  {\bibfnamefont {T.~W.}\ \bibnamefont {H{\''}ansch}},\ }\href@noop {}
  {\bibfield  {journal} {\bibinfo  {journal} {Applied Physics B}\ }\textbf
  {\bibinfo {volume} {89}},\ \bibinfo {pages} {535} (\bibinfo {year}
  {2007})}\BibitemShut {NoStop}%
\bibitem [{\citenamefont {Quraishi}\ \emph {et~al.}(2014)\citenamefont
  {Quraishi}, \citenamefont {Diddams},\ and\ \citenamefont
  {Hollberg}}]{Quraishi2014}%
  \BibitemOpen
  \bibfield  {author} {\bibinfo {author} {\bibfnamefont {Q.}~\bibnamefont
  {Quraishi}}, \bibinfo {author} {\bibfnamefont {S.~A.}\ \bibnamefont
  {Diddams}}, \ and\ \bibinfo {author} {\bibfnamefont {L.}~\bibnamefont
  {Hollberg}},\ }\href@noop {} {\bibfield  {journal} {\bibinfo  {journal}
  {Optics Communications}\ }\textbf {\bibinfo {volume} {320}},\ \bibinfo
  {pages} {84} (\bibinfo {year} {2014})}\BibitemShut {NoStop}%
\bibitem [{\citenamefont {Coluccelli}\ \emph {et~al.}(2015)\citenamefont
  {Coluccelli}, \citenamefont {Cassinerio}, \citenamefont {Gambetta},
  \citenamefont {Laporta},\ and\ \citenamefont {Galzerano}}]{Coluccellii2015}%
  \BibitemOpen
  \bibfield  {author} {\bibinfo {author} {\bibfnamefont {N.}~\bibnamefont
  {Coluccelli}}, \bibinfo {author} {\bibfnamefont {M.}~\bibnamefont
  {Cassinerio}}, \bibinfo {author} {\bibfnamefont {A.}~\bibnamefont
  {Gambetta}}, \bibinfo {author} {\bibfnamefont {P.}~\bibnamefont {Laporta}}, \
  and\ \bibinfo {author} {\bibfnamefont {G.}~\bibnamefont {Galzerano}},\
  }\href@noop {} {\bibfield  {journal} {\bibinfo  {journal} {Scientific
  Reports}\ }\textbf {\bibinfo {volume} {5}},\ \bibinfo {pages} {16338}
  (\bibinfo {year} {2015})}\BibitemShut {NoStop}%
\bibitem [{\citenamefont {Thiel}\ \emph {et~al.}(2018)\citenamefont {Thiel},
  \citenamefont {Roslund}, \citenamefont {De}, \citenamefont {Fabre},\ and\
  \citenamefont {Treps}}]{Val2018}%
  \BibitemOpen
  \bibfield  {author} {\bibinfo {author} {\bibfnamefont {V.}~\bibnamefont
  {Thiel}}, \bibinfo {author} {\bibfnamefont {J.}~\bibnamefont {Roslund}},
  \bibinfo {author} {\bibfnamefont {S.}~\bibnamefont {De}}, \bibinfo {author}
  {\bibfnamefont {C.}~\bibnamefont {Fabre}}, \ and\ \bibinfo {author}
  {\bibfnamefont {N.}~\bibnamefont {Treps}},\ }\href@noop {} {\bibfield
  {journal} {\bibinfo  {journal} {arXiv}\ }\textbf {\bibinfo {volume}
  {physics.ins-det/1806.02198}} (\bibinfo {year} {2018})}\BibitemShut {NoStop}%
\bibitem [{\citenamefont {Schmeissner}\ \emph
  {et~al.}(2014{\natexlab{a}})\citenamefont {Schmeissner}, \citenamefont
  {Roslund}, \citenamefont {Fabre},\ and\ \citenamefont
  {Treps}}]{TheSchmeissnerPaper}%
  \BibitemOpen
  \bibfield  {author} {\bibinfo {author} {\bibfnamefont {R.}~\bibnamefont
  {Schmeissner}}, \bibinfo {author} {\bibfnamefont {J.}~\bibnamefont
  {Roslund}}, \bibinfo {author} {\bibfnamefont {C.}~\bibnamefont {Fabre}}, \
  and\ \bibinfo {author} {\bibfnamefont {N.}~\bibnamefont {Treps}},\
  }\href@noop {} {\bibfield  {journal} {\bibinfo  {journal} {Physical Review
  Letters}\ }\textbf {\bibinfo {volume} {113}},\ \bibinfo {pages} {263906}
  (\bibinfo {year} {2014}{\natexlab{a}})}\BibitemShut {NoStop}%
\bibitem [{\citenamefont {Grynberg}\ \emph {et~al.}(2010)\citenamefont
  {Grynberg}, \citenamefont {Aspect},\ and\ \citenamefont
  {Fabre}}]{grynberg2010}%
  \BibitemOpen
  \bibfield  {author} {\bibinfo {author} {\bibfnamefont {G.}~\bibnamefont
  {Grynberg}}, \bibinfo {author} {\bibfnamefont {A.}~\bibnamefont {Aspect}}, \
  and\ \bibinfo {author} {\bibfnamefont {C.}~\bibnamefont {Fabre}},\
  }\href@noop {} {\emph {\bibinfo {title} {Introduction to quantum optics: from
  the semi-classical approach to quantized light}}}\ (\bibinfo  {publisher}
  {Cambridge University Press},\ \bibinfo {year} {2010})\BibitemShut {NoStop}%
\bibitem [{\citenamefont {Jian}\ \emph {et~al.}(2012)\citenamefont {Jian},
  \citenamefont {Pinel}, \citenamefont {Fabre}, \citenamefont {Lamine},\ and\
  \citenamefont {Treps}}]{jian2012real}%
  \BibitemOpen
  \bibfield  {author} {\bibinfo {author} {\bibfnamefont {P.}~\bibnamefont
  {Jian}}, \bibinfo {author} {\bibfnamefont {O.}~\bibnamefont {Pinel}},
  \bibinfo {author} {\bibfnamefont {C.}~\bibnamefont {Fabre}}, \bibinfo
  {author} {\bibfnamefont {B.}~\bibnamefont {Lamine}}, \ and\ \bibinfo {author}
  {\bibfnamefont {N.}~\bibnamefont {Treps}},\ }\href@noop {} {\bibfield
  {journal} {\bibinfo  {journal} {Optics Express}\ }\textbf {\bibinfo {volume}
  {20}},\ \bibinfo {pages} {27133} (\bibinfo {year} {2012})}\BibitemShut
  {NoStop}%
\bibitem [{\citenamefont {Schmeissner}\ \emph
  {et~al.}(2014{\natexlab{b}})\citenamefont {Schmeissner}, \citenamefont
  {Thiel}, \citenamefont {Jacquard}, \citenamefont {Fabre},\ and\ \citenamefont
  {Treps}}]{TheSchmeissnerPaperAboutCavities}%
  \BibitemOpen
  \bibfield  {author} {\bibinfo {author} {\bibfnamefont {R.}~\bibnamefont
  {Schmeissner}}, \bibinfo {author} {\bibfnamefont {V.}~\bibnamefont {Thiel}},
  \bibinfo {author} {\bibfnamefont {C.}~\bibnamefont {Jacquard}}, \bibinfo
  {author} {\bibfnamefont {C.}~\bibnamefont {Fabre}}, \ and\ \bibinfo {author}
  {\bibfnamefont {N.}~\bibnamefont {Treps}},\ }\href@noop {} {\bibfield
  {journal} {\bibinfo  {journal} {Optics Letters}\ }\textbf {\bibinfo {volume}
  {39}},\ \bibinfo {pages} {3603} (\bibinfo {year}
  {2014}{\natexlab{b}})}\BibitemShut {NoStop}%
\bibitem [{\citenamefont {Holman}\ \emph {et~al.}(2003)\citenamefont {Holman},
  \citenamefont {Jones}, \citenamefont {Marian}, \citenamefont {Cundiff},\ and\
  \citenamefont {Ye}}]{holman2003detailed}%
  \BibitemOpen
  \bibfield  {author} {\bibinfo {author} {\bibfnamefont {K.~W.}\ \bibnamefont
  {Holman}}, \bibinfo {author} {\bibfnamefont {J.~R.}\ \bibnamefont {Jones}},
  \bibinfo {author} {\bibfnamefont {A.}~\bibnamefont {Marian}}, \bibinfo
  {author} {\bibfnamefont {S.~T.}\ \bibnamefont {Cundiff}}, \ and\ \bibinfo
  {author} {\bibfnamefont {J.}~\bibnamefont {Ye}},\ }\href@noop {} {\bibfield
  {journal} {\bibinfo  {journal} {IEEE Journal of Selected Topics in Quantum
  Electronics}\ }\textbf {\bibinfo {volume} {9}},\ \bibinfo {pages} {1018}
  (\bibinfo {year} {2003})}\BibitemShut {NoStop}%
\bibitem [{\citenamefont {Helbing}\ \emph {et~al.}(2002)\citenamefont
  {Helbing}, \citenamefont {Steinmeyer}, \citenamefont {Keller}, \citenamefont
  {Windeler}, \citenamefont {Stenger},\ and\ \citenamefont
  {Telle}}]{Helbing2002}%
  \BibitemOpen
  \bibfield  {author} {\bibinfo {author} {\bibfnamefont {F.~W.}\ \bibnamefont
  {Helbing}}, \bibinfo {author} {\bibfnamefont {G.}~\bibnamefont {Steinmeyer}},
  \bibinfo {author} {\bibfnamefont {U.}~\bibnamefont {Keller}}, \bibinfo
  {author} {\bibfnamefont {R.~S.}\ \bibnamefont {Windeler}}, \bibinfo {author}
  {\bibfnamefont {J.}~\bibnamefont {Stenger}}, \ and\ \bibinfo {author}
  {\bibfnamefont {H.~R.}\ \bibnamefont {Telle}},\ }\href@noop {} {\bibfield
  {journal} {\bibinfo  {journal} {Optics Letters}\ }\textbf {\bibinfo {volume}
  {27}},\ \bibinfo {pages} {194} (\bibinfo {year} {2002})}\BibitemShut
  {NoStop}%
\bibitem [{\citenamefont {Newbury}\ and\ \citenamefont
  {Swann}(2007)}]{Newbury2007}%
  \BibitemOpen
  \bibfield  {author} {\bibinfo {author} {\bibfnamefont {N.~R.}\ \bibnamefont
  {Newbury}}\ and\ \bibinfo {author} {\bibfnamefont {W.~C.}\ \bibnamefont
  {Swann}},\ }\href@noop {} {\bibfield  {journal} {\bibinfo  {journal} {Journal
  of Optical Society of America B}\ }\textbf {\bibinfo {volume} {24}},\
  \bibinfo {pages} {1756} (\bibinfo {year} {2007})}\BibitemShut {NoStop}%
\bibitem [{\citenamefont {Herr}\ \emph {et~al.}(2014)\citenamefont {Herr},
  \citenamefont {Brasch}, \citenamefont {Jost}, \citenamefont {Mirgorodskiy},
  \citenamefont {Lihachev}, \citenamefont {Gorodetsky},\ and\ \citenamefont
  {Kippenberg}}]{Herr2014}%
  \BibitemOpen
  \bibfield  {author} {\bibinfo {author} {\bibfnamefont {T.}~\bibnamefont
  {Herr}}, \bibinfo {author} {\bibfnamefont {V.}~\bibnamefont {Brasch}},
  \bibinfo {author} {\bibfnamefont {J.}~\bibnamefont {Jost}}, \bibinfo {author}
  {\bibfnamefont {I.}~\bibnamefont {Mirgorodskiy}}, \bibinfo {author}
  {\bibfnamefont {G.}~\bibnamefont {Lihachev}}, \bibinfo {author}
  {\bibfnamefont {M.}~\bibnamefont {Gorodetsky}}, \ and\ \bibinfo {author}
  {\bibfnamefont {T.~J.}\ \bibnamefont {Kippenberg}},\ }\href@noop {}
  {\bibfield  {journal} {\bibinfo  {journal} {Physical Review Letters}\
  }\textbf {\bibinfo {volume} {113}},\ \bibinfo {pages} {123901} (\bibinfo
  {year} {2014})}\BibitemShut {NoStop}%
\end{thebibliography}%
\end{document}